\documentclass{elsart}

\usepackage{amssymb}
\usepackage{graphicx}
\begin{document}

\begin{frontmatter}



\title{Optimization of the Ballistic Guide Design for the SNS FNPB 8.9
\AA\ Neutron Line}


\author{Takeyasu ~M.~Ito\corauthref{cor1}\thanksref{label1}}
\corauth[cor1]{Corresponding author}
\ead{ito@lanl.gov}
\thanks[label1]{Present address: Los Alamos National Laboratory, Los
  Alamos, NM 87545}
\author{Christopher B. Crawford},
\author{Geoffrey ~L.~Greene}
\address{The Department of Physics and Astronomy, The University of
  Tennessee\\
Knoxville, TN 37996\\
and\\ Physics Division, Oak Ridge National Laboratory, Oak Ridge, TN, 37831}

\begin{abstract}
The optimization of the ballistic guide design for the SNS Fundamental
Neutron Physics Beamline 8.9 \AA\ line is described. With a careful
tuning of the shape of the curve for the tapered section and the width
of the straight section, this optimization resulted in more than a 75\%
increase in the neutron flux exiting the 33~m long guide over a
straight $m=3.5$ guide with the same length.
\end{abstract}

\end{frontmatter}

\section{Introduction}
Recent years have seen a remarkable progress in the technology of
producing supermirrors, both in terms of the critical angle and the
reflectivity, which resulted in substantial increase in neutron flux
at various neutron beam facilities. However, for transporting neutrons
over long distances ($\gtrsim 30$~m), the finite reflectivity for
angles above $\theta_c$ (the critical angle for natural nickel) causes
a significant loss in the neutron flux exiting the guide. (The
reflectivity of a good $m=3.5$ guide is 80\% at $m=3.5$).

The so-called ``ballistic guide'' geometry, first proposed by
Mezei~\cite{MEZ97}, allows neutrons to be transported over long
distances without significant losses. The first ballistic guide using
supermirror was installed as the H113 beamline at
ILL~\cite{HAE02}. This guide was 72~m long. It diverged linearly in
width from 6~cm to 9~cm over the first 10~m. The width stayed at 9~cm
over the next 44~m, and then converged back linearly to 6~cm. It
showed an excellent performance providing a factor four increase in
neutron fluence over the $^{58}$Ni coated straight guide installed
before. In Ref.~\cite{SCH04}, an optimization of a 46.8~m long
ballistic guide geometry is presented and the results are compared to
the existing straight guide 1RNR14 at SINQ at PSI. In this study,
parabolic and elliptic tapering in both horizontal and vertical
directions was considered for the diverging and converging horns as
well as straight (=linear) tapering. It was found that for their input
neutron distribution, a guide geometry with elliptically tapered
sections performs better than a ballistic guide geometry with straight
taper, giving a factor 5 higher neutron flux compared to that of the
existing $m=2$ straight guide.

We have performed a careful optimization of the ballistic guide design
for the SNS FNPB 8.9 \AA\ Neutron Beam Line using a custom made Monte
Carlo simulation program. We carefully optimized various aspects of
the ballistic guide geometry. In particular, we adopted a new
parameterization of the shape of the tapered section, which allowed us
to optimize the shape of the tapered section in an efficient
manner. With a careful tuning of the shape of the tapered section and
the width of the straight section, this optimization resulted in more
than a 75\% increase in the neutron flux exiting the 33~m long guide
over that of a straight $m=3.5$ guide with the same
length\footnote{The gain factor of 1.75 should not be directly
compared to the gain factors of 4 and 5 obtained in
Refs.~\cite{HAE02} and \cite{SCH04} because of the differences in the length of
the guide, the divergence of the beam, and the coating on the straight
guide that the ballistic guide is compared to.}. In this paper, we
describe the optimization process and present the obtained
results. One important point that needs to be emphasized is that the
optimum choice for various aspects of the ballistic guide geometry,
such as the shape of the tapered section and the widths of the
straight section, strongly depends on the angular and positional
distribution of the neutron flux entering the ballistic guide.

This paper is structured as follows. In
section~\ref{sec:general_consideration}, the principle of the
ballistic guide and some important design considerations are
reviewed. Then in section~\ref{section:optimization}, after a brief
description of the SNS FNPB beamline and its characteristics, the
optimization process including the Monte Carlo program is described in
detail, and the results are presented.

\section{General Consideration}
\label{sec:general_consideration}
\subsection{Ballistic guide -- the principle and some design considerations}
\label{sec:principle}
\begin{figure}
\begin{center}
\includegraphics[bb=200 60 400 780,angle=90, width=10cm]{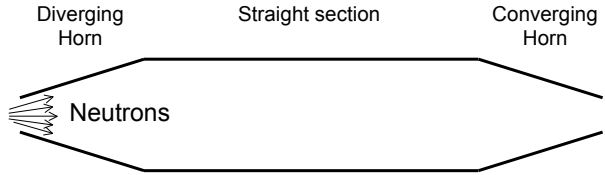}
\end{center}
\caption{Schematic of a ballistic guide\label{fig:ballistic_schematic}}
\end{figure}

The term ``ballistic guide'' refers to an arrangement in which a 
diverging horn is followed by a wide (and usually long) straight
section, and then by a converging horn (see
Fig.~\ref{fig:ballistic_schematic}). With such an
arrangement, the divergence (angular spread) of the incident neutron
distribution is turned to a spatial spread by the diverging horn,
which results in a reduced loss during the propagation through the
straight section because of the smaller angles (i.e. higher
reflectivity) and fewer bounces (due to the smaller angles and the
larger guide width).

Liouville's theorem states that the following inequality holds for the
phase space occupied by an ensemble of neutrons traveling through a
ballistic guide:
\begin{equation}
\label{eq:phase_space_rot}
\Delta \theta_i \cdot \Delta x_i
\leq \Delta \theta_S \cdot \Delta x_S,
\end{equation}
and where $\Delta \theta_i$ and $\Delta x_i$ are the angular and
spatial spreads of the initial distribution of the neutrons entering
the diverging horn of the ballistic guide, and $\Delta \theta_S$ and
$\Delta x_S$ are those for the neutrons in the straight section right
after the diverging horn. A similar inequality holds for the neutrons
in the straight section right before the converging horn and the
neutrons exiting the converging horn. Note, however, in general in
both the diverging and converging horns, there is an unavoidable loss
of neutron flux due to the fact that the reflectivity is less than
unity for angles larger than $\theta_c$. In fact, as we will see
later, the loss in the diverging horn is a major source of loss in
ballistic guide geometry.

The task of optimizing the design is to perform this ``phase space
rotation'' (from large $\Delta \theta_i$ and small $\Delta x_i$ to
small $\Delta \theta_S$ and large $\Delta x_S$) in the diverging horn
as efficiently as possible, i.e. as closely to the equality limit as
possible, while minimizing the loss (due to the finite reflectivity)
in the diverging horn itself.  An inefficient phase space rotation
would result in a larger than necessary guide width for the straight
section, which would not only be more expensive to build but would
also cause a larger loss in the converging horn. Also, the shape of
the converging horn has to be optimized to minimize additional
losses during the second phase space rotation.

\begin{figure}
\begin{center}
\includegraphics[width=5in]{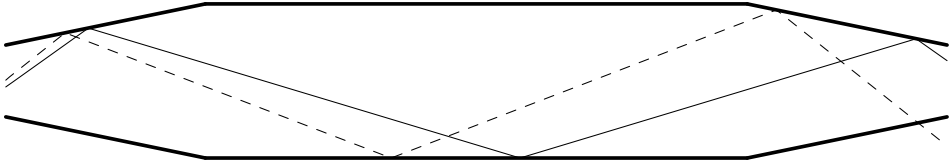}
\end{center}
\caption{ Illustration of how the converging horn does not do the
  exact reverse of the diverging horn because of the non point-source
  nature of the distribution (see main
  text). \label{fig:neutron_tracks}}
\end{figure}

The converging horn does not exactly reverse the effect of the
diverging horn on the neutron phase space. Rather, a loss of neutrons
also occurs in the converging horn, due to the nonlinear nature of
ballistic transport.  The loss in the converging horn is another major
contributor to the loss in the ballistic guide. This point is
illustrated in Fig.~\ref{fig:neutron_tracks}, where the solid and
dashed lines represent the trajectories of two neutrons that had
slightly different initial conditions (the neutrons travel from left
to right). Despite the small difference in the initial conditions, one
(solid line) makes it to the end whereas the other (dashed line) does
not (these trajectories were calculated by the simulation program
described in this document). The neutron represented by the dashed
trajectory gets lost when it makes the second interaction with the
wall of the converging horn, where the angle was too large for the
neutron to be reflected (the first interaction in the converging horn
has the same angle as that in the diverging horn). It can also be seen
that the neutron with the dashed trajectory too would make it to the end
without making the second interaction with the wall in the converging
horn, if the straight section were a little shorter or the neutron
started at a slightly different location.

For a typical supermirror, the reflectivity is virtually unity for
angles less than $\theta_c$. Therefore, $\Delta \theta_S$ does not
have to be infinitesimally small. It should rather be around
$\theta_c$. Otherwise $\Delta x_S$ would be unnecessarily large,
which would cause a larger loss in the converging horn. This argument
provides a good estimate for the optimum guide width of the straight
section. In general, if the input neutron distribution has an angular
spread that extends to $m_i\times \theta_c$, then, the
optimum guide width is expected to be in the neighborhood of $m_i$
times the width of the entrance of the diverging horn (or perhaps
slightly narrower, because the narrower the guide is in the straight
section, the smaller the loss is in the converging horn).

\subsection{Curved taper}
\label{sec:curved}
\begin{figure}
\begin{center}
\includegraphics[bb=150 100 420 750, angle=90, width=4in]{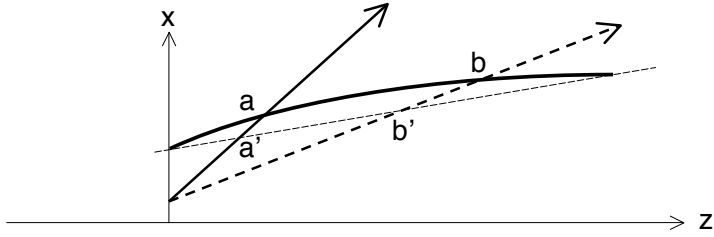}
\end{center}
\caption{Comparison between a curved and a straight taper. The curved
  taper is represented by the thick solid curve. The thin straight
  dashed line represents the straight taper. The solid and dashed arrows
  represent neutrons coming in at large and small angles,
  respectively. \label{fig:curved1}}
\end{figure}

The simplest realization of a ballistic guide uses a straight taper
for the diverging and converging horns. However, a more efficient
phase space rotation of Eq.~(\ref{eq:phase_space_rot}) can be achieved
by using a curved taper as depicted in Fig.~\ref{fig:curved1}. This
figure~\ref{fig:curved1} illustrates how the use of a curved taper for
the diverging horn can result in a smaller loss in the horn and in a
more favorable neutron angular distribution in the straight
section. The solid arrow represents a trajectory of a neutron with a
large angle. It would hit the wall of the horn at a larger angle at
point {\tt a'} for the straight taper than it would at point {\tt a}
for the curved taper. Since the reflectivity is larger for smaller
angles of incident (when the angle is larger than $\theta_c$), the
curved taper causes a smaller loss for large angle neutrons than the
straight taper. The dashed arrow represents a trajectory of a neutron
with a small angle. In this case, the neutron would hit the wall at a
larger angle for the curved taper than it would for the straight
taper. However, in this case, the angle is already small (smaller than
$\theta_c$) so hitting the wall at a larger angle does not lead to a
loss. Instead, the final angle resulting from the bounce is closer to
being parallel to the $z$ axis for the curved taper, which results in
a reduced loss in the converging horn. (Note if all the neutrons have
a trajectory parallel to the $z$ axis, the loss in the converging horn
is substantially reduced.)

\begin{figure}
\begin{center}
\includegraphics[bb=150 100 430 760,angle=90, width=4in]{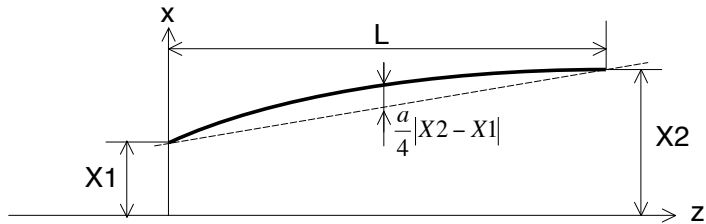}
\end{center}
\caption{Parameterization used for curved tapers in this
  study (see the main text). \label{fig:curved2}}
\end{figure}

We therefore considered using a curved taper for the diverging and
converging horns in this study. As is obvious from the discussion in
the previous paragraph, the shape of the curved taper has to be
optimized for the given neutron distribution. To conveniently
parameterize the shape of the curve, we used the following expression
for the shape of curved tapers:
\begin{equation}
\label{eq:curve}
x = \frac{(X2-X1)}{L} z + X1 - a\frac{z-L}{L}\frac{z}{L}|X2-X1|,
\end{equation}
where $X1$ and $X2$ are the half width of the entrance and exit of the
horn and $L$ is the length of the horn, as shown in
Fig.~\ref{fig:curved2}. The neutron beam travels in the positive $z$
direction in this figure, and $x$ is perpendicular to $z$. Note that
the first two terms if Eq.~(\ref{eq:curve}) give the straight line
that connects the two points determined by the entrance and exit
widths and the length of the horn. The third term is a quadratic in
$z$, fixed at the endpoints, which represents the first order
deviation from the straight line. The dimensionless parameter $a$
determines the size of this term: $a=0$ corresponds to the straight
line, and $a=1$ corresponds to matching the derivative at the end of
the horn with the straight section. It gives a curve that deviates
from the straight line by $\frac{|X2-X1|}{4}$ at $z=L/2$ (note that
both the factor $z$ and $z-L$ are divided by $L$ so that $a$ is
dimensionless).

There is no point in using a parabola, since the source is not
point-like. The advantage of using this parameterization, instead of
using parabola, hyperbola, or ellipsoid, is that it is linear in the
limit of $a=0$ instead of in the asymptotic limit of some parameter
$p\rightarrow \infty$. No matter which shape we use -- parabola,
hyperbola, or ellipsoid -- the actual deviation from the straight line
will be small because the horn is several meters long whereas it is
only a few tens of centimeters wide. Therefore our parameterization
should capture the essence of the curved guide. In other words, we do
not expect the results to be too dependent of the details of the
shape, which was confirmed for our particular neutron distribution by
adding terms in higher order in $z$ and seeing no significant gain in
performance (see Section~\ref{sec:fulloptimization}).

\section{Optimization of the FNPB 8.9 \AA\ beamline}
\label{section:optimization}
\subsection{SNS and FNPB}
\begin{figure}
\begin{center}
\includegraphics[bb=40 120 570 670, width=12cm]{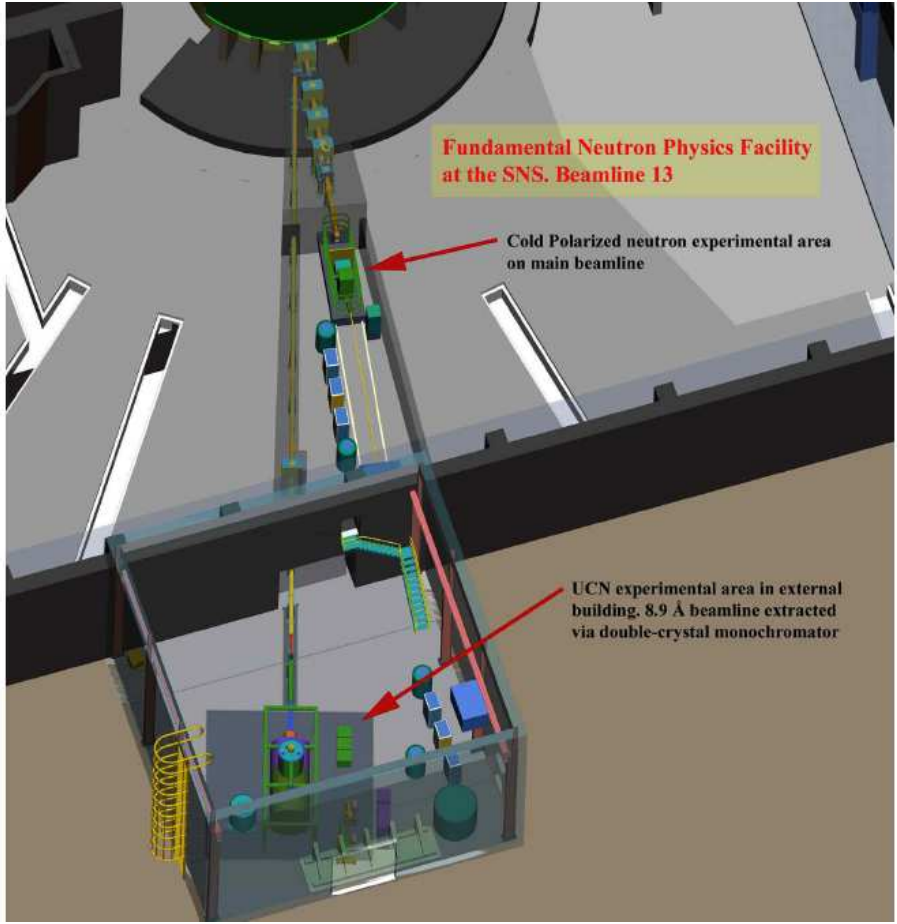}
\end{center}
\caption{Schematic of the layout of the Fundamental Neutron Physics
  Facility at the SNS.\label{fig:FNPB}}
\end{figure}

The Spallation Neutron Source (SNS), currently under construction at
the Oak Ridge National Laboratory, is an accelerator-based neutron
source, and will provide the world's most intense pulsed neutron beams
for scientific research and industrial development~\cite{SNS}.  The
Fundamental Neutron Physics Beamline (FNPB), one of the 24 neutron
beamlines in the SNS target hall, is dedicated to fundamental physics
using cold and ultracold neutrons. Figure~\ref{fig:FNPB} shows the
schematic of the layout of the FNPB beamline.
The FNPB has two neutron beamlines, the ``Cold Neutron Line'' and 
the 8.9\AA\ Line (or ``UCN Line''). The 8.9\AA\ Line is dedicated
to experiments that will uses the superthermal process in superfluid
liquid helium to produce ultra-cold neutrons. The 8.9\AA\ neutrons
will be selected by a double crystal monochromator and will be sent to
an external building located about 30~m downstream. 

A calculation~\cite{HUF05} shows that a neutron fluence of $0.94\times
10^9$~n/s/\AA\ can be obtained when a straight $m=3.5$ guide with a
cross section of 12~cm~$\times$~14~cm and a length of 33~m is used to
transport the 8.9\AA\ neutrons from the monochromator to the external
building. The goal of this study is to find the optimum geometry for a
ballistic guide which transports the 8.9\AA\ neutrons from the
monochromator to the external building.

\subsection{Characteristics of the incident neutron distribution}
\label{sec:neutron_dist}
\begin{figure}
\begin{center}
\includegraphics[width=5in]{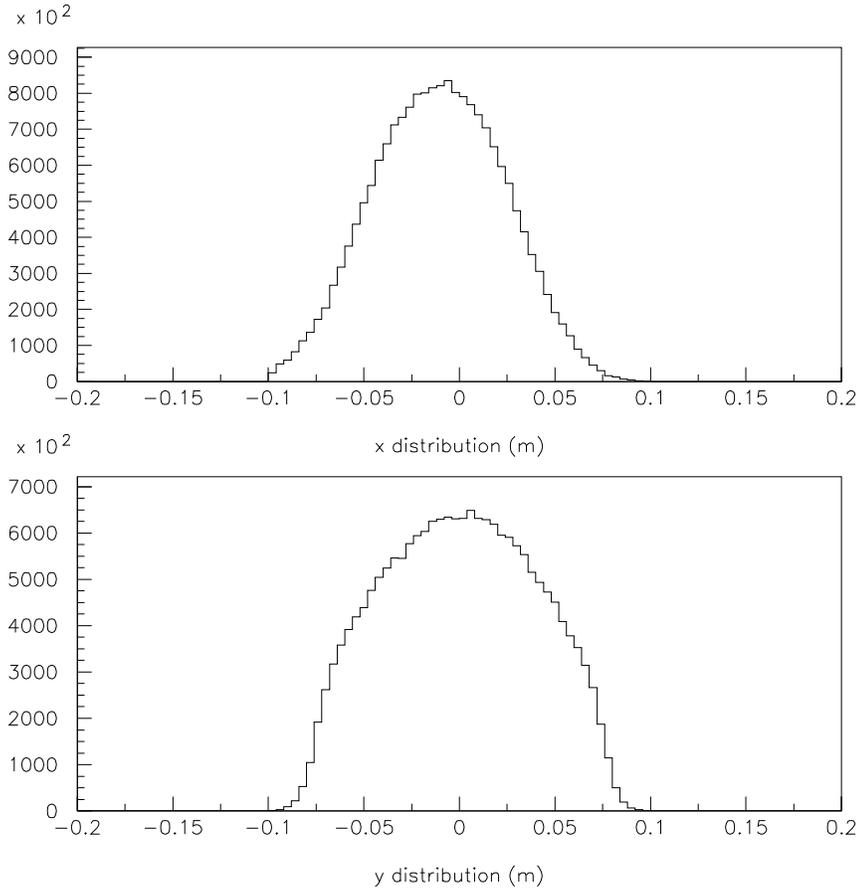}
\end{center}
\caption{Horizontal (top) and vertical (bottom) position distributions
  of the neutrons coming out of the double crystal
  monochromator\label{fig:pos_in}}
\end{figure}

\begin{figure}
\begin{center}
\includegraphics[width=5in]{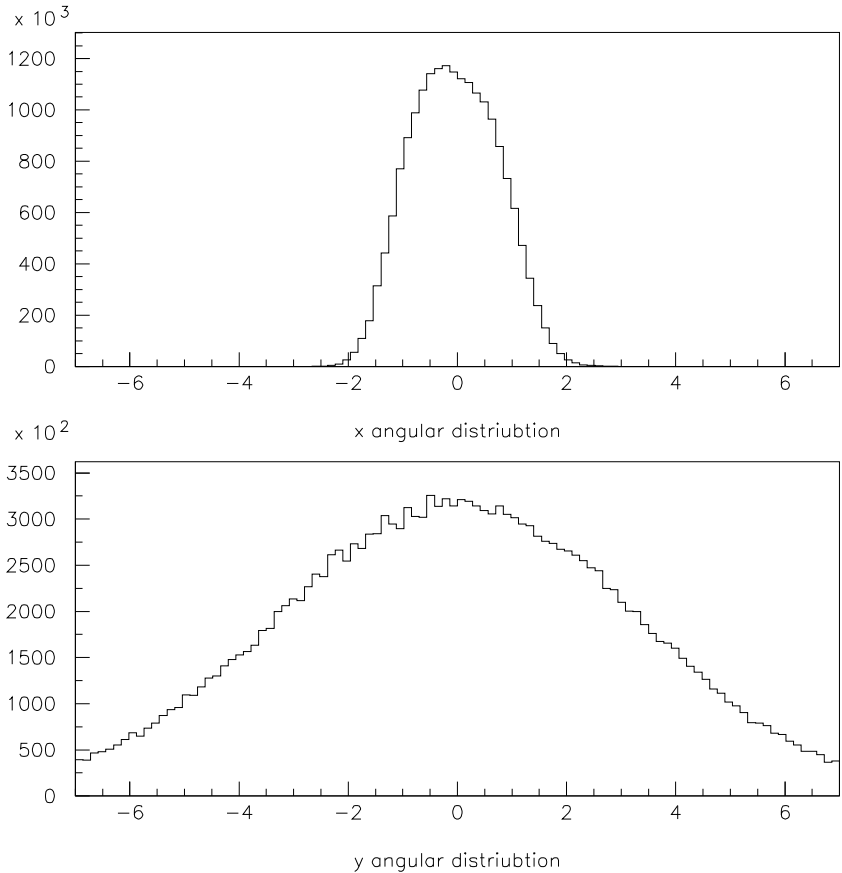}
\end{center}
\caption{Horizontal (top) and vertical (bottom) angular distributions
  of the neutrons coming out of the double crystal monochromator. The
  angles are in the unit of $\theta_c$.\label{fig:ang_in}}
\end{figure}

As mentioned in Section~\ref{sec:general_consideration}, the optimum choice
for various aspect of the ballistic guide geometry, such as the shape
of the tapered section and the widths of the straight section,
strongly depends on the angular and position distribution of the
neutron flux entering the ballistic guide.  It is therefore important
to examine the characteristics of the incident neutron distribution in
order to intelligently optimize the design of the ballistic
guide. Figures~\ref{fig:pos_in} and \ref{fig:ang_in} show the position
and angular distribution in the horizontal ($x$) and vertical ($y$)
directions of the neutrons coming out of the double crystal
monochromator. These distributions were obtained from a Monte Carlo
program written by P.~Huffman based on {\sc McStas}~\cite{MCSTAS},
which generates neutrons according to the neutron input source files
for the SNS target and tracks neutrons through the FNPB beamline
elements including the double crystal monochromator that extracts the
8.9 \AA\ neutrons~\cite{HUF05}.

The position distributions shown in Fig.~\ref{fig:pos_in} confirm
that the choice of the cross section of 12~cm~$\times$~14~cm for the
entrance of the ballistic guide is reasonable.

For the angular distributions, from Fig.~\ref{fig:ang_in} we notice
that the horizontal angular spread is only $|\Delta \theta_x|<
2\theta_c$ whereas the vertical angular spread is as large as $|\Delta
\theta_y| \sim 4\theta_c$. This is because of the mosaic structure of
the crystals that are used in the double crystal monochromator, and
also because of how the two crystals are arranged in the double crystal
monochromator. The angular spread caused by the randomly oriented
small ``blocks'' in the mosaic of one crystal cancels that of the
other crystal to the first order in the horizontal direction but not
in the vertical direction when the two crystals are arranged
horizontally. The narrow horizontal angular distribution also
indicates that the neutron guide upstream of the double crystal
monochromator is long enough for the neutrons with large angles be be
lost while they travel from the SNS target to the monochromator.

The implication of these two angular distributions to the design of
the ballistic guide is as follows (see
Fig.~\ref{fig:ballistic_schematic2} for the notation):
\begin{itemize}
\item For the horizontal direction, from
  Eq.~(\ref{eq:phase_space_rot}) and the succeeding argument, the
  optimum width of the straight section is expected to be $XW2 \sim
  2\times XW1 = 24$~cm.
\item Also, since the angular spread is relatively small in the
  horizontal direction, it is expected that the supermirror coating 
  can have a smaller value for $m$ on the side walls of the diverging
  horn.
\item On the other hand, in the vertical direction, because of the
  large angular spread, the width of the straight section should
  ideally be $YW2 \sim 4\times YW1 =
  56$~cm. However, it is mechanically limited to $\sim 30$~cm.
\end{itemize}

As discussed in Section~\ref{sec:simulation}, we varied $XW2$ between 15~cm
and 30~cm in search of the optimum ballistic guide design.

We note that the angular spread could be reduced by using graphite
crystals with a smaller mosaic angle~\cite{HUF05b}.
\subsection{Simulation program}
\label{sec:simulation}
The optimization of the ballistic guide was performed using a Monte
Carlo simulation. A new neutron ray tracing package, dubbed {\sc
neutrack 8.9}, was written to perform a series of transmission
calculations to optimize the ballistic guide for the 8.9 \AA\
line. While there was already an {\sc McStas} program to simulate the
whole FNPB beamline system, we decided that it was advantageous to
develop a new ray tracing package and write a whole new program using
it to design the 8.9 \AA\ line for several reasons, including the
following:
\begin{itemize}
\item The {\sc McStas} FNPB simulation tracks neutrons all the way
  from the SNS target down to the end of the beam line for all
  wavelengths. The fraction of the neutrons that get selected by the
  monochromator and get sent to the 8.9 \AA\ line is very
  small. However, optimizing the ballistic guide design only involve
  simulating neutrons downstream of the monochromator. The simulation
  can be done more efficiently if the program simulates only the
  components downstream of the monochromator, generating only 8.9 \AA\
  neutrons at the entrance to the ballistic guide.
\item Information important in designing a ballistic guide -- such as,
  where exactly in the system neutrons are lost and what the
  distribution is like for the angle at which the neutrons interact
  with each segment of supermirror -- is not easily accessible by the
  user in {\sc McStas}.
\end{itemize}

{\sc neutrack 8.9} is designed to simulate only a single wavelength of
neutrons. The user has a complete access to information such as where
exactly in the system neutrons are lost and what the distribution is
like for the angle at which the neutrons interact with each segment of
supermirror.

A Monte Carlo simulation program using {\sc neutrack 8.9} was written
to simulate a ballistic guide for the 8.9 \AA\ line. Neutrons were
generated so that their distribution reproduces that of the neutrons
coming out of the double crystal monochromator calculated by the {\sc
  McStas} FNPB simulation. The neutrons were then introduced into the
ballistic guide, and the behavior of the neutrons in the guide was
simulated.  The performance of various geometries was evaluated using
the transmission ({\sc txmit}) -- the ratio of the number of
neutrons exiting the guide to the number of neutrons entering
the guide --  and the relative transmission (rel. {\sc txmit}) --
the ratio of the transmission of the guide geometry under
consideration to that of the 12~cm $\times$ 14~cm straight
guide. (Note that since the input to the ballistic guide is the same
for all different geometries, the relative transmission is the same as
the output neutron flux or fluence normalized to that of the straight
guide.)

Various parameters including properties of the supermirrors were
adjusted so that the results of this 8.9 \AA\ line simulation program
agreed with those from the {\sc McStas} FNPB simulation program for
the same ballistic guide geometry. It was confirmed for several
selected geometries that the results for the relative transmission
from both simulation programs for the same geometry agreed within the
statistical error of the simulation ($\sim 1$\%) as shown in
Table~\ref{tab:comparison}. It was important to reproduce the input
neutron distributions obtained from the {\sc McStas} FNPB simulation
in the 8.9 \AA\ line simulation, including the correlation between the
vertical position and the vertical angle to obtain an agreement
between the two simulation programs.  Because of the simplicity of
{\sc neutrack 8.9} and because the 8.9 \AA\ line simulation program
only simulates the 8.9 \AA\ line downstream of the monochromator, it
is a more than two orders of magnitude faster than the {\sc McStas}
FNPB simulation program to obtain results with the same statistical
precision.
\begin{table}
\caption{Comparison between {\sc McStas} and {\sc neutrack
    8.9}. [Description of the guide geometries (for the parameter
    definition, see Table~\protect\ref{tab:parameters}):
    a) straight guide with a 12~cm~$\times$~14~cm cross section,
    $m=3.5$ for all four walls;  b) ballistic guide with straight
    taper with $L1=L3=9$~m, $XW2=30$~cm, $m1_{TB}=m1_{LR}=3.5$,
    $m2_{TB}=m2_{LR}=2.0$, $m3_{TB}=m3_{LR}=3.5$; c) ballistic guide
    with curved taper with $L1=L3=9$~m, $XW2=30$~cm,
    $m1_{TB}=m1_{LR}=3.5$, $m2_{TB}=m2_{LR}=2.0$,
    $m3_{TB}=m3_{LR}=3.5$, $a_H=a_V=1.0$; d) ballistic guide with
    curved taper with $L1=L3=7$~m, $XW2=20$~cm, $m1_{TB}=m1_{LR}=3.5$,
    $m2_{TB}=m2_{LR}=2.0$, $m3_{TB}=m3_{LR}=3.5$, $a_H=0.3$,
    $a_V=1.1$.]  }
\label{tab:comparison}
\begin{center}
\begin{tabular}{ccccc}\\ \hline\hline
Guide & \multicolumn{2}{c}{{\sc McStas}} 
& \multicolumn{2}{c}{{\sc neutrack 8.9}}\\
 & {\sc txmit} & rel. {\sc txmit} & {\sc txmit} & rel. {\sc txmit} \\ \hline
a) Straight guide& 0.3854(24) & 1.000 & 0.3976(5)
& 1.000 \\
b) Ballistic, straight taper & 0.5827(31) & 1.512(12) & 0.6031(5) &
1.517(2) \\
c) Ballistic, curved taper & 0.6375(32) & 1.654(13) & 0.6611(5) &
1.663(2) \\
d) Ballistic, curved taper, narrow & 0.6759(33) & 1.754(14) & 0.6976(5) &
1.755(2) \\ \hline\hline
\end{tabular}
\end{center} 
\end{table}

\subsection{Physics-motivated optimization}
\label{sec:physoptimization}
Monte Carlo simulations were performed for various ballistic guide
geometries to find the optimum geometry [high performance (= high
transmission) and low cost]. We first performed an optimization in
which the ranges of various parameters were selected based on the
considerations given in Section~\ref{sec:general_consideration}.

\begin{figure}
\begin{center}
\includegraphics[bb=120 30 500 800,angle=90, width=5in]{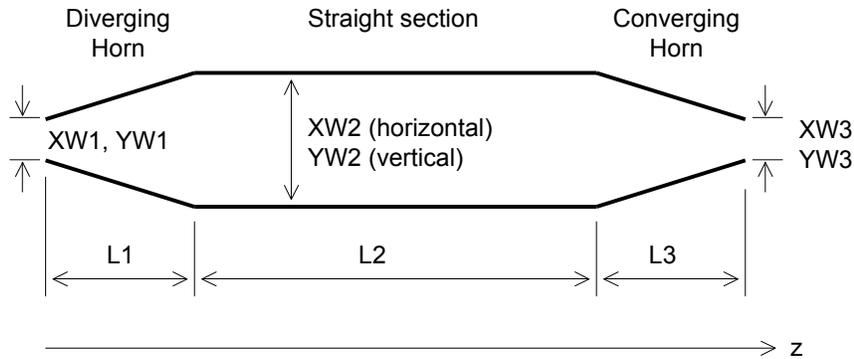}
\end{center}
\caption{Schematic of a ballistic guide with parameters used to
  describe the geometry\label{fig:ballistic_schematic2}}
\end{figure}

The parameters used to describe the ballistic guide geometry are
listed in Table~\ref{tab:parameters} along with the value for the
fixed parameters or the range of variation for optimized parameters.
Some of the parameters are illustrated in
Fig.~\ref{fig:ballistic_schematic2}. The values and the ranges for the
parameters reflect various boundary conditions (e.g. the largest guide
width practically possible is $\sim 30$~cm) and considerations (such
as those discussed in Sections~\ref{sec:principle} and
\ref{sec:neutron_dist}). For this study the length of the converging
horn $L3$ was set to be the same as the length of the diverging horn
$L1$ since no significant gain was seen when $L3$ was varied
independent of $L1$ in a preliminary study.

When performing Monte Carlo simulation for geometries with curved
taper, the curved taper was approximated by a series of short segments
with straight tapers at different angles. Usually, each horn was
divided into five segments to keep the computation time short (the
computation time scales approximately as the number of the mirror
segments). For a few selected geometries, a finer division was also
tried. Results from five segments agreed with results from fourteen
segments (for $L1=7$~m this corresponds to 50~cm long segments) within
1\%.

\begin{table}
\caption{Parameters used to describe the ballistic guide geometry. The
  taper curvature parameters $a_H$ and $a_V$ are common for the diverging
  and converging horns.}
\label{tab:parameters}
\begin{center}
\begin{tabular}{cll}\\ \hline\hline
Parameter & \multicolumn{1}{c}{Description} 
& \multicolumn{1}{c}{Value} \\ \hline
$L1$ & length of the diverging horn & 5, 6, 7, 8, 9~m \\
$L2$ & length of the straight section & 33~m $- (L1+L3)$ \\
$L3$ & length of the converging horn & $L3=L1$ \\
$XW1$ & width of the entrance & 12~cm (fixed) \\
$XW2$ & width of the straight section & 15~cm $-$ 30~cm  \\
$XW3$ & width of the exit & 12~cm (fixed) \\
$YW1$ & height of the entrance & 14~cm (fixed) \\
$YW2$ & height of the straight section & 30~cm (fixed)  \\
$YW3$ & height of the exit & 14~cm (fixed) \\
$a_H$ & curvature parameter for the horns (left and right walls) & $0-1.5$ \\
$a_V$ & curvature parameter for the horns (top and bottom walls) & $0-1.5$ \\
$m1_{TB}$ & $m$ for the diverging horn, top and bottom walls & 3.5\\
$m1_{LR}$ & $m$ for the diverging horn, left and right walls & 2.0, 3.5\\
$m2_{TB}$ & $m$ for the straight section, top and bottom walls & 2.0\\
$m2_{LR}$ & $m$ for the straight section, left and right walls & 1.5, 2.0\\
$m3_{TB}$ & $m$ for the converging horn, top and bottom walls & 3.5\\
$m3_{LR}$ & $m$ for the converging horn, left and right walls & 2.0, 3.5\\ \hline\hline
\end{tabular}
\end{center}
\end{table}

\begin{figure}[p]
\begin{center}
\includegraphics[width=5in]{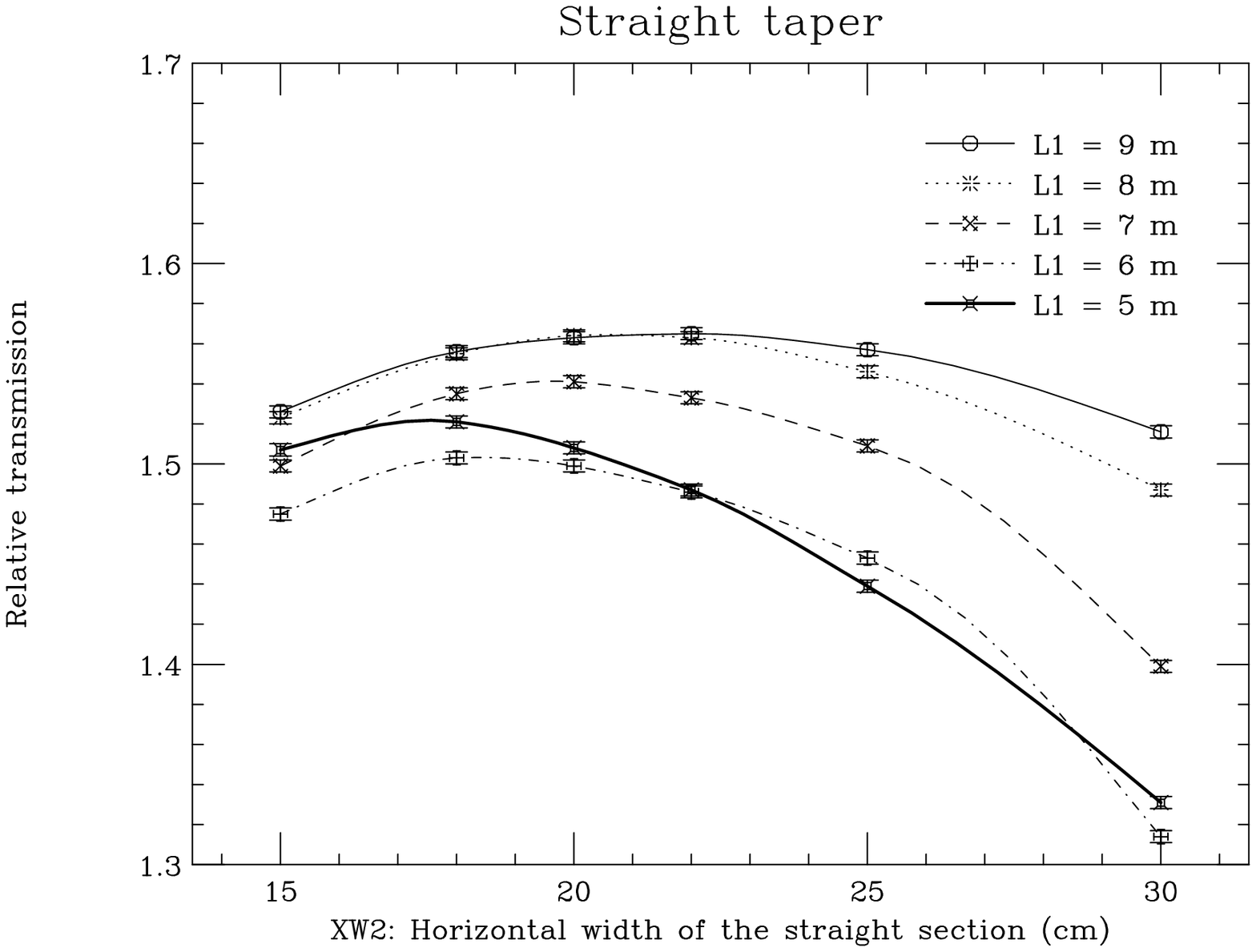}
\end{center}
\caption{Relative transmission vs horizontal width of the straight
  section for geometries with straight taper. \label{fig:neutron_vs_xw2_s}}
\end{figure}

\begin{figure}[p]
\begin{center}
\includegraphics[width=5in]{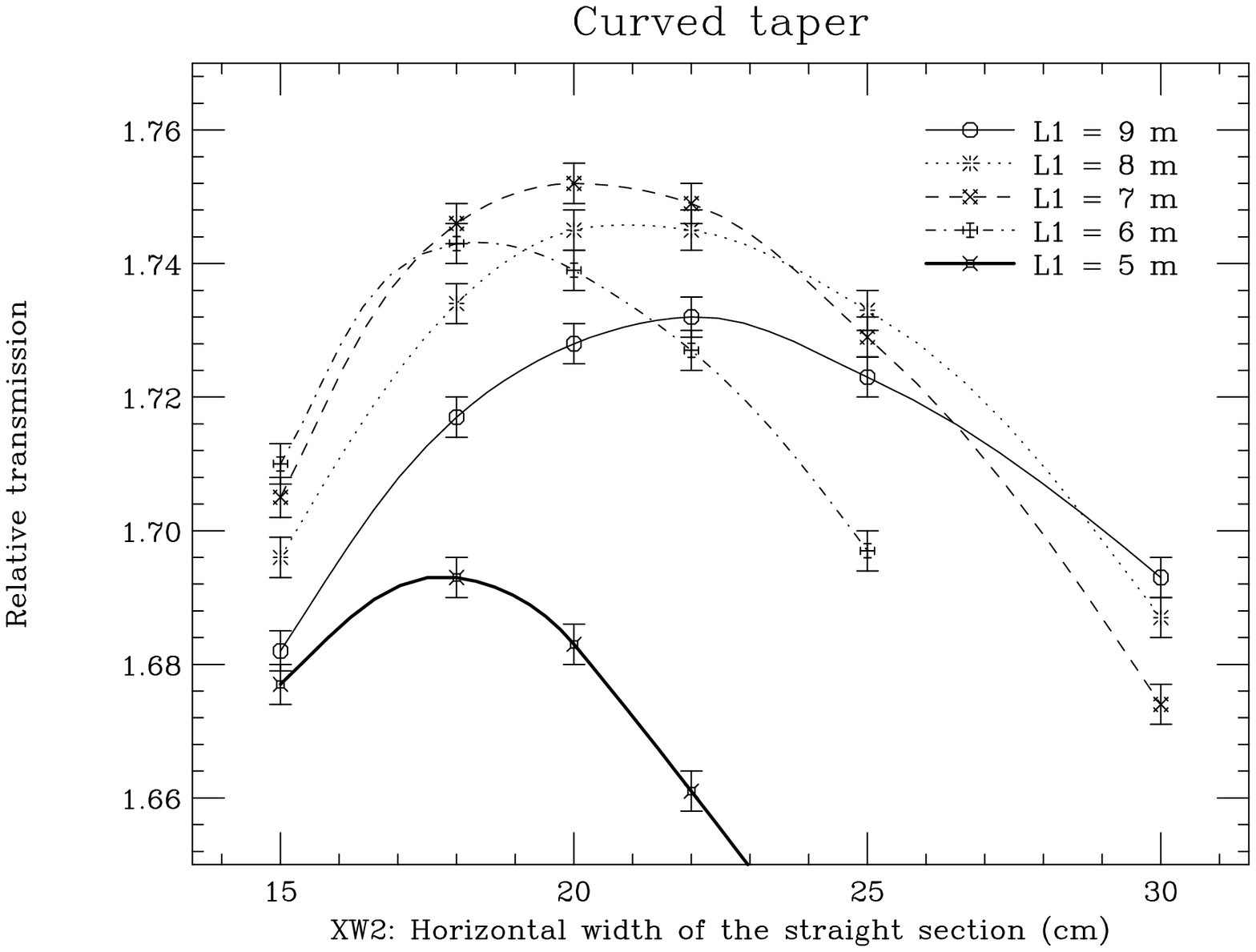}
\end{center}
\caption{Relative transmission vs horizontal width of the straight
  section for geometries with curved taper. \label{fig:neutron_vs_xw2_c}}
\end{figure}

The dependence of the relative transmission on $L1$ (=$L3$) and $XW2$
is shown in Fig.~\ref{fig:neutron_vs_xw2_s} for geometries with
straight taper and in Fig.~\ref{fig:neutron_vs_xw2_c} for geometries
with curved taper.
For curved taper, the curvature parameters $a_H$ and $a_V$ were
optimized for each combination of $L1$ and $XW2$. We found that the
optimum values of $a_H$ and $a_V$ do not have a strong dependence on
$XW2$, but do depend on the value of $L1$. The optimum values of $a_H$
and $a_V$ for different values of $L1$ are listed in
Table~\ref{tab:optimum_a}.
\begin{table}[t]
\caption{Optimum values for $a_H$ and $a_V$ for different values of
  $L1$}
\label{tab:optimum_a}
\begin{center}
\begin{tabular}{ccc} \\ \hline\hline
$L1$ & $a_H$ & $a_V$ \\ \hline
5~m & 0.4 & 0.7 \\
6~m & 0.4 & 0.9 \\
7~m & 0.3 & 1.1 \\
8~m & 0.6 & 1.3 \\
9~m & 0.4 & 1.4 \\ \hline\hline
\end{tabular}
\end{center}
\end{table}

\begin{figure}
\begin{center}
\includegraphics[width=4in]{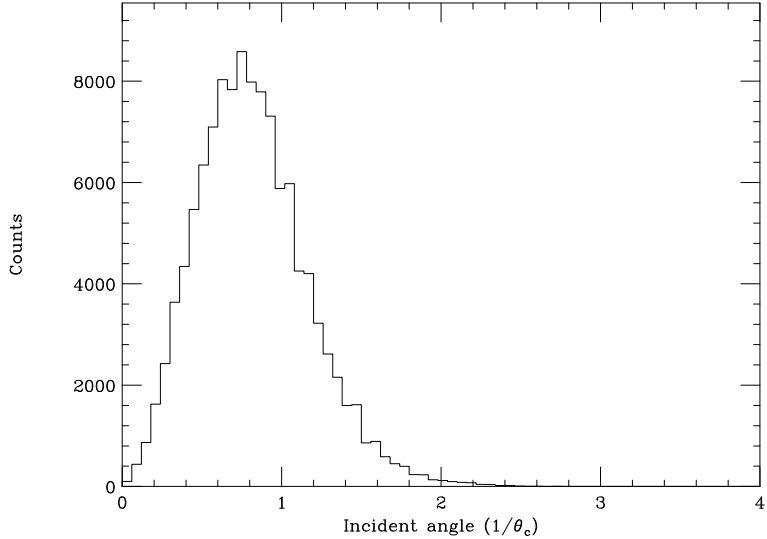}
\end{center}
\caption{Distribution of the incident angle of the neutrons on the
  side wall of the diverging horn. (For this plot, a geometry with
  $L1=8$~m and $XW2=20$~cm with straight taper was used as an
  example.) \label{fig:ang_horn_side_wall}}
\end{figure}

\begin{figure}
\begin{center}
\includegraphics[width=12cm]{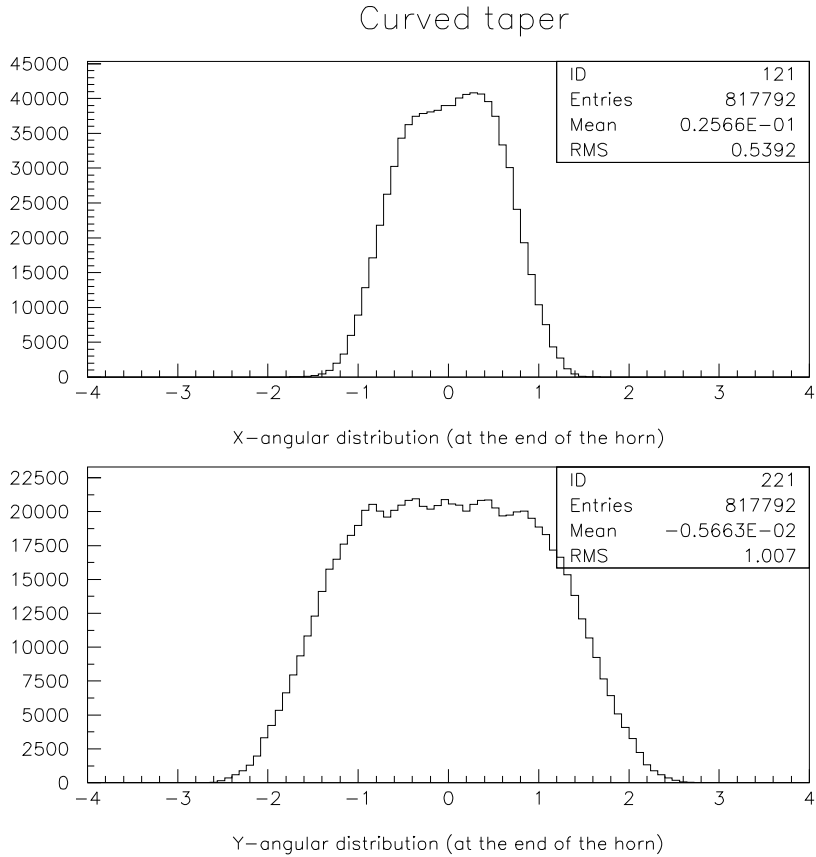}
\end{center}
\caption{Horizontal (top) and vertical (bottom) position distributions
  of the neutrons in the ballistic guide right after the diverging
  horn with curved taper. The angles are in the unit of $\theta_c$.
  (For this plot, a geometry with $L1=7$~m, $XW2=20$~cm,
  $m1_{TB}=3.5$, $m1_{LR}=2.0$ was used as an example.)
  \label{fig:ang_after_horn_c}}
\end{figure}

For all the results plotted in these two Figures, the following values
of $m$ were used for the supermirror coating: $m1_{TB}=3.5$,
$m1_{LR}=2.0$, $m2_{TB}=2.0$, $m2_{LR}=1.5$, $m3_{TB}=3.5$,
$m3_{LR}=3.5$. We now discuss how each of these parameters was
optimized. Because of the large vertical angular spread of the initial
neutron distribution, it is important to use as high an $m$ as
possible for the top and bottom walls of the diverging horn
($m1_{TB}=3.5$). However, because of the rather narrow horizontal
divergence of the initial neutron distribution, it is not necessary to
use as high an $m$ for the side walls. In fact, reducing $m1_{LR}$
down to 2.0 from 3.5 did not lead to any reduction of the output
neutron flux. This can also be seen from
Fig.~\ref{fig:ang_horn_side_wall}, which shows a distribution of the
incident angle of the neutrons on the side wall of the diverging
horn. It is seen that there are hardly any neutrons incident on the
side wall of the diverging horn with an angle larger than
$2\times\theta_c$.
For the straight section, the values for the $m$ were determined to be
$m2_{TB}=2.0$ and $m2_{LR}=1.5$ by looking at the the horizontal and
vertical angular distributions of the neutrons right after the
diverging horn, which are shown in Fig.~\ref{fig:ang_after_horn_c}. We
further confirmed the validity of these choices by varying the values
of these two $m$'s and seeing negligible increase in the output
neutron flux when the values of the $m$'s were increased and seeing a
significant reduction in the output neutron flux when the values of
the $m$'s were decreased.  For the converging horn, based on the
discussion given in Section~\ref{sec:principle}, it is important to
use the largest possible value of $m$. Therefore, we decided to use
$m3_{TB}=3.5$, $m3_{LR}=3.5$. It was confirmed by actual Monte Carlo
calculations that lowering these $m$'s significantly decreases the
output neutron flux. Reducing the $m$
for the side walls of the diverging horn and the straight section from
$m1_{LR}=3.5$ and $m2_{LR}=2.0$ to $m1_{LR}=2.0$ and $m2_{LR}=1.5$
reduced the guide price by $\sim 150,000$ USD or more, which is a
substantial cost saving.
\begin{figure}
\begin{center}
\includegraphics[width=12cm]{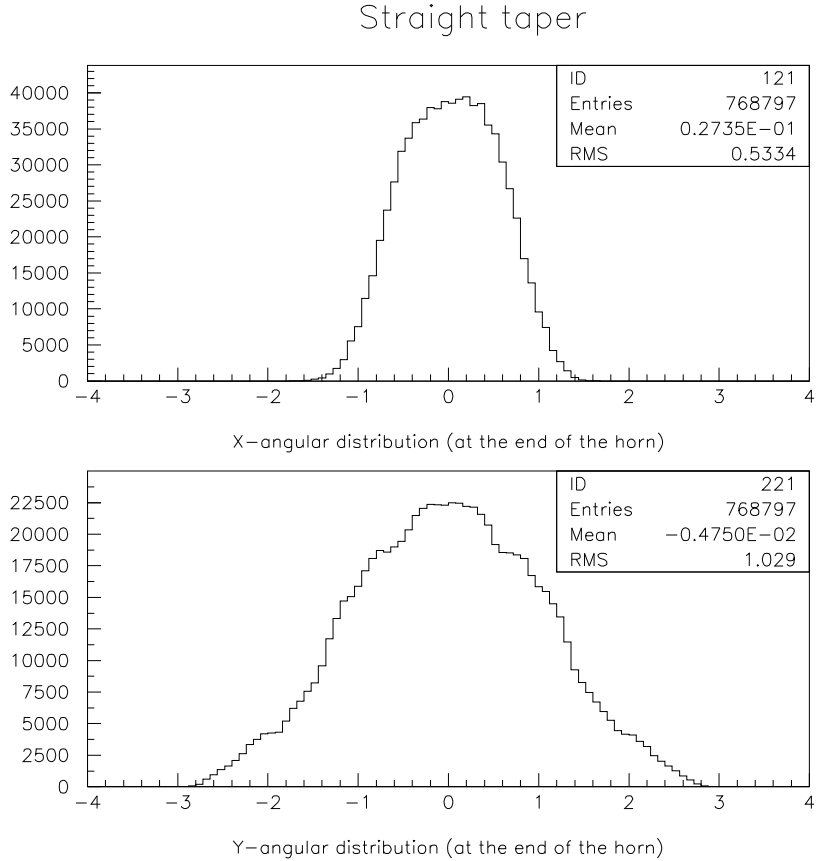}
\end{center}
\caption{Horizontal (top) and vertical (bottom) angular distributions
  of the neutrons in the ballistic guide right after the diverging
  horn straight taper. The angles are in units of $\theta_c$. (For
  this plot, a geometry with $L1=7$~m, $XW2=20$~cm, $m1_{TB}=3.5$,
  $m1_{LR}=2.0$ was used as an example.)
  \label{fig:ang_after_horn_s}}
\end{figure}

Figure~\ref{fig:neutron_vs_xw2_s} shows that geometries with straight
taper favor a large $L1$, with $L1=8$~m and $L1=9$~m being the
optimum. On the other hand, Figure~\ref{fig:neutron_vs_xw2_c} shows
that geometries with curved taper favor a shorter $L1$, with $L1=7$~m
giving the maximum relative transmission (with $XW2=20$~cm).

To see the effect of curved taper, let us compare
Fig.~\ref{fig:ang_after_horn_c} and
Fig.~\ref{fig:ang_after_horn_s}. These Figures contain the horizontal
and vertical angular distributions of neutrons exiting the diverging
horn for a guide geometry with curved taper
(Fig.~\ref{fig:ang_after_horn_c}) and straight taper
(Fig.~\ref{fig:ang_after_horn_s}). Geometries with $L1=7$~m and
$XW2=20$~cm are used here as examples. Although for the horizontal
angular distribution the curved taper does not make a big difference
because of the rather narrow initial distribution, it makes an obvious
difference for the vertical distribution. Both of the two effects that
were discussed in Section~\ref{sec:curved} can be seen, i.e., a
smaller loss in the diverging horn, and a more favorable angular
distribution after the diverging horn. The transmission through the
diverging horn is 82\% for curved taper, compared to 77\% for the
straight taper and the difference in the vertical angular distribution
is obvious.

Coming back to Figs.~\ref{fig:neutron_vs_xw2_s} and
\ref{fig:neutron_vs_xw2_c}, from these figures we observe the
following three points:
\begin{itemize}
\item By using a ballistic guide we obtain a substantial gain in
  neutron flux ($\sim 55$\% for straight taper and $\sim 75$\% for
  curved taper as compared to the straight guide).
\item Ballistic guides with curved taper give higher a output neutron
  flux than ballistic guides with straight taper.
\item The optimum horizontal width for the straight section is around
  $20$~cm (slightly narrower than expected based on a naive argument
  given in Section~\ref{sec:principle}). 
\end{itemize}

In Table~\ref{tab:lostneutron}, the fraction of neutrons lost in
various places in the ballistic guide are listed for the optimum
geometry with curved taper and for the optimum geometry with straight
taper. As expected, the use of the curved taper reduces the loss in
the diverging and converging horns.
\begin{table}
\caption{Fractions of neutrons lost in various places in the ballistic
guide for the optimum geometry with curved taper and for the optimum
geometry with straight taper}
\label{tab:lostneutron}
\begin{center}
\begin{tabular}{lcc}\\ \hline\hline
Description & Curved taper & Straight taper\\ \hline
Lost in the diverging horn (left or right walls) & 0.005 & 0.006\\
Lost in the diverging horn (top or bottom walls) & 0.177 & 0.243\\
Lost in the straight section (left or right walls) & 0.008 & 0.007\\
Lost in the straight section (top or bottom walls) & 0.052 & 0.049\\
Lost in the converging horn (left or right walls) & 0.008 & 0.008\\
Lost in the converging horn (top or bottom walls) & 0.053 & 0.066\\
Survived & 0.697 & 0.622\\ \hline\hline
\end{tabular} 
\end{center}
\end{table}

\subsection{Full optimization}
\label{sec:fulloptimization}
In order to verify the results of the physics motivated optimization,
we proceeded to do a full-scale optimization of the ballistic guide
geometry, fitting for as many degrees of freedom as possible.  Two
goals of this optimization were a) to investigate the effect of the
shape of the diverging and converging horns, and b) to verify that the
two horns should indeed be symmetric.

To investigate the shape of the horn, a cubic term was added to
Eq.~\ref{eq:curve}:
\begin{equation}
  \label{eq:cubic}
  \hat x = \hat z\left[1 + a(1-\hat z)\left(1+b(1-\hat z)\right)\right],
\end{equation}
in normalized coordinates $0<\hat x<1$ and $0<\hat z<1$, where $x = X_1 +
|X_2-X_1|\,\hat x$ and $z = L_1 + |L_2-L_1|\,\hat z$.  The coordinates
$(X_1,L_1)$ are at the entrance/exit of the horn, while $(X_2,L_2)$ connect to
the straight section.  The cubic term with coefficient $b$ was added in such a
way that it reduces to Eq.~\ref{eq:curve} in the limit $b=0$, to a straight
line in the limit $a=0$, and has the same slope $m=1-a$ at the end of the horn
($z=L_2$).  By imposing the constraint of no inflection points over the length
of the horn, the cubic term is limited to the range $-{\textstyle \frac12} < b
< 1$.  This extra term allows one to match up the slope of the horn with the
straight section ($a=1$) while simultaneously adjusting the slope at $z=0$
within the range ${\textstyle\frac32} < d\hat x/d\hat z < 3$.

Noting the large difference between $a_H$ and $a_V$ in
Table~\ref{tab:optimum_a}, a second parameter $c=L_V/L_H-1$ was added, relaxing
the curved horn to have different lengths in the horizontal and vertical
sides.  The purpose of this additional parameter was to ensure that the
shape of the horn in the vertical direction was not constrained by its
horizontal length.  Separate parameters $a_H$, $a_V$, $b_H$, $b_V$, and $c$
were optimized for both the diverging and converging horns.

The multiparameter optimization was done using a custom computer code
{\sc minoise} based on the conjugate gradient technique
(Ref.\cite{nr}, sec. 10.6).  The line minimization algorithm was
modified to handle uncertainty in the minimization function (the
transmission) due to statistical errors associated with the Monte
Carlo technique.  Instead of bracketing and searching for the minimum
by golden means, the transmission is sampled at $n$ points along the
line, and fit for the parabolic minimum.  The algorithm also has the
capability to: a) narrow the range if $\chi^2$ of the fit is too
large; b) zoom out if uncertainty in the minimum is too large or the
parabola has the wrong curvature; c) recenter the range on the
minimum, adding extra points to fill in the gaps; and d) increase the
density of points for the final fit after finding the range.  Because
of the cost of each Monte Carlo simulation, care was taken to reuse
old points and only do new simulations where necessary.  In order to
perform the $400-1500$ Monte Carlo simulations necessary for a full
parameter optimization, the {\sc neutrack 8.9} based ballistic guide
simulation described in section~\ref{sec:simulation} was executed from
an MPI~\cite{mpich} wrapper, and run in parallel on a 40 CPU farm. In
this manner, simulations of 200000 events could be run in less than
5~s each.  

Three optimizations were carried out, listed in
Table~\ref{tab:optimum_multi}.  The first optimization was done to
find the optimal height of the guide, $YW2=37.6$~cm.  This is very
close to the limit of 30~cm, and the transmission changed very little
to due this constraint.  The second optimization verified the values
of $L$ and $a$ determined in the physics-motivated section, and showed
that the guide should be reasonably symmetric along its length.  The
third fit, a full-parameter optimization, improved very little over
the previously obtained transmission.  This supports the above
arguments that the guide performance is insensitive to the details of
its curvature, and shows that the guide is well-optimized for the
angular distribution of neutrons at the source.  In fact, very little
transmission was gained by increasing the $m$-value of the supermirros
in the diverging horn.

\begin{table}[t]
  \caption{Multiparameter optimizations of transmission: line 1) fit of optimal
    guide height, 2) verification of conventional parameters, 3) fit for extra
    curvature parameters.}
  \label{tab:optimum_multi}
  \begin{center}
    \begin{tabular}{l@{~~}c@{~}c@{~}c@{~}c@{~}c@{~}c@{~~}r@{~}l@{~~}c@{~}c@{~}c@{~}c@{~}c@{~}c@{~}c}
      \hline\hline
      &\multicolumn{6}{l}{diverging horn} &
      \multicolumn{2}{c}{straight} &
      \multicolumn{6}{l}{converging horn} & {\sc txmit} \\
      &$L1$ & $a_H$ & $a_V$ & $b_H$ & $b_V$ & $c$ & $XW$ & $YW$ &
      $L3$ & $a_H$ & $a_V$ & $b_H$ & $b_V$ & $c$ &  \\ \hline
      1) & 8.8 & .22 & 1.14 & --- & --- & --- & .218 & .376 & 9.3 & .58 & 1.25& --- & --- & --- & 74.4 \\ 
      2) & 7.2 & .29 & 1.16 & --- & --- & --- & .202 & ---  & 7.4 & .61 & 1.10& --- & --- & --- & 70.2 \\ 
      3) & 8.4 & .20 & 1.16 &-.15 & .53 &-.056& .206 & ---  & 8.2 & .57 & 1.12& .49 & .17 &-.038& 70.5 \\ 
      \hline\hline
    \end{tabular}
  \end{center}
\end{table}

\section{Summary}
The design for the ballistic guide for the FNPB 8.9 \AA\ neutron line
was optimized using Monte Carlo simulation. A simulation program
written using {\sc neutrack 8.9}, a neutron ray tracing package
specially developed for this purpose. It was shown that it is
possible to increase the output neutron flux (and fluence) by more
than 70\% (as compared to that obtained by the straight guide) by
using a ballistic guide with curved taper. With a properly designed
ballistic guide, a neutron fluence of $1.65\times 10^9$~n/s/\AA\ is
expected, which should be compared to $0.94\times 10^9$~n/s/\AA, a
fluence expected for the straight guide.

\section{Acknowledgments}
We thank P.~Huffman for valuable suggestions. This work was
supported in part by US Department of Energy Division of Nuclear
Physics through grant number DE-FG02-03ER41258.

\end{document}